\documentclass{PoS}

\title{Using Drell-Yan $A_{FB}$ to constrain PDFs}
\ShortTitle{Using $A_{FB}$ to constrain Parton Distribution Functions}

\author{\speaker{A. Bodek}, J. Y. Han, A. Khukhunaishvili, W. Sakumoto
\\    Department of Physics and Astronomy, University of Rochester, Rochester, NY 14627, USA\\
        E-mail: \email{bodek@pas.rochester.edu}}

\abstract{
We show that measurements of the  forward-backward charge asymmetry ($A_{FB}(M,y)$)  of Drell-Yan  dilepton events produced at hadron colliders 
provide a new powerful tool to  constrain  Parton Distribution Functions (PDFs).  PDF uncertainties are the dominant source of  systematic error in precision measurements at hadron colliders (e.g.  $\sin^2\theta_{eff}(M_Z)$, $\sin^2\theta_{W}=1-M_W^2/M_Z^2$ and the mass of the W boson).  We show that the
   $\chi^2$  values of fits to extract  $\sin^2\theta_{eff}^{lept}(M_Z)$  from  $A_{FB}(M,y)$ with different PDF replicas can be used to place additional constraints on PDFs. In turn, using these constrained PDFs  significantly reduces the PDF errors in precision measurements  of electroweak parameters.  The measurement
   of the on-shell $\sin^2\theta_{W}=1-M_W^2/M_Z^2$ is equivalent to an indirect measurement of the W mass. The  errors in this
   indirect measurement of the W mass are competitive with direct measurements. For
   example, with 200 fb$^{-1}$ at 13 TeV, the expected error in the indirect measurement of
   the W mass is $\pm$9 MeV.
          }

\FullConference{The XXIII International Workshop on Deep Inelastic Scattering and Related Subjects\\
		April 27 - May 1, 2015\\
                Southern Methodist University\\
		Dallas, Texas 75275}

\begin{document}

     Within the standard model, measurements of the mass of the $Z$ boson and top quark, in combination with the
    mass of the Higgs boson,  can be used to predict  the mass of the W boson.
   At present,  the average of  all direct measurements of the mass of the W boson
    (80385$\pm$15 MeV) is about one standard deviation higher than the prediction of the standard
    model.  Predictions of  supersymmetric models
     for the W mass are also higher than the predictions of the standard model. 
     Therefore, more precise measurements of the mass of the W boson
    are of great  interest.
     
     Alternatively, the W mass can also be extracted indirectly from
     measurements of the on-shell electroweak mixing angle $\sin^2\theta_{W}$ 
    by using the  relation $\sin^2\theta_{W}=1-M_W^2/M_Z^2$.   
      Measurements of the  forward-backward charge asymmetry in Drell-Yan
         dilepton events produced at hadron colliders 
(in the region of the $Z$ pole)   have been used to measure
 the value of the  {\it effective} electroweak (EW)  mixing
  angle $\sin^2\theta_{eff}^{lept} (M_Z)$\cite {cdf-ee,cdf-mumu,dzero,ATLAS}.
   In addition, by incorporating electroweak radiative corrections in the analysis
   the CDF collaboration has also measured the  {\it on-shell} EW mixing angle  $\sin^2\theta_W$\cite{cdf-ee,cdf-mumu}.
    An error of $\pm$0.00030 in the measurement of $\sin^2\theta_{W}$ 
      is equivalent to an indirect measurement  of the W mass to a precision of $\pm$15 MeV.  However,  the PDF
      error  quoted in the most recent measurement of $\sin^2\theta_{eff}$ by the ATLAS collaboration\cite{ATLAS}
      at the LHC is $\pm$0.00090.  Therefore, a significant reduction in the PDF errors is needed. Here, we 
      show how $A_{FB}$  data (both at the Tevatron and LHC)  also provide a new powerful tool  to constrain  PDFs.  Addition details
      can be found in Ref. \cite{bodek-afb}.

\begin{figure}[hb]
\includegraphics[width=7.5 cm, height=6.0 cm]{{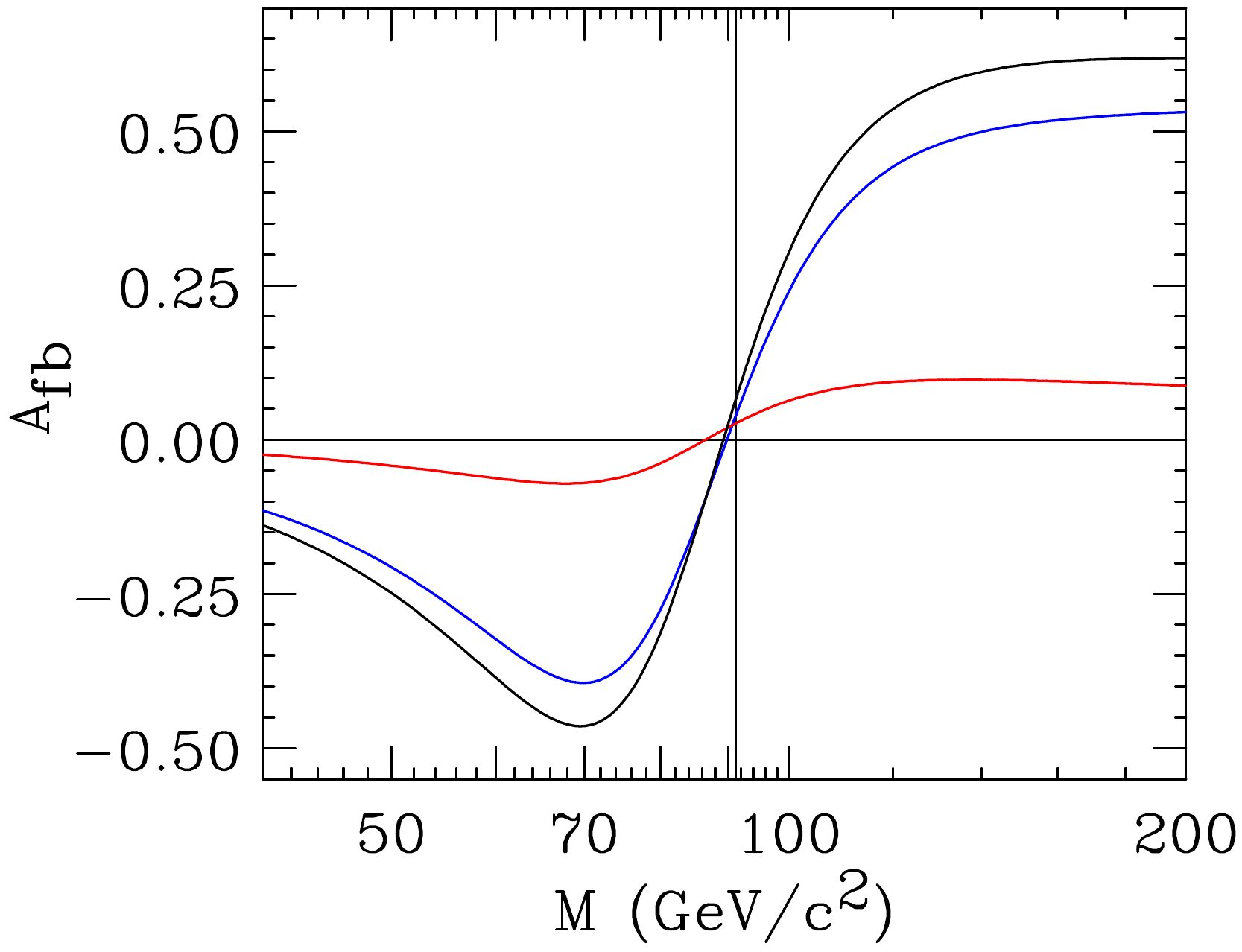}}
\includegraphics[width=7.5 cm, height=7.0 cm]{{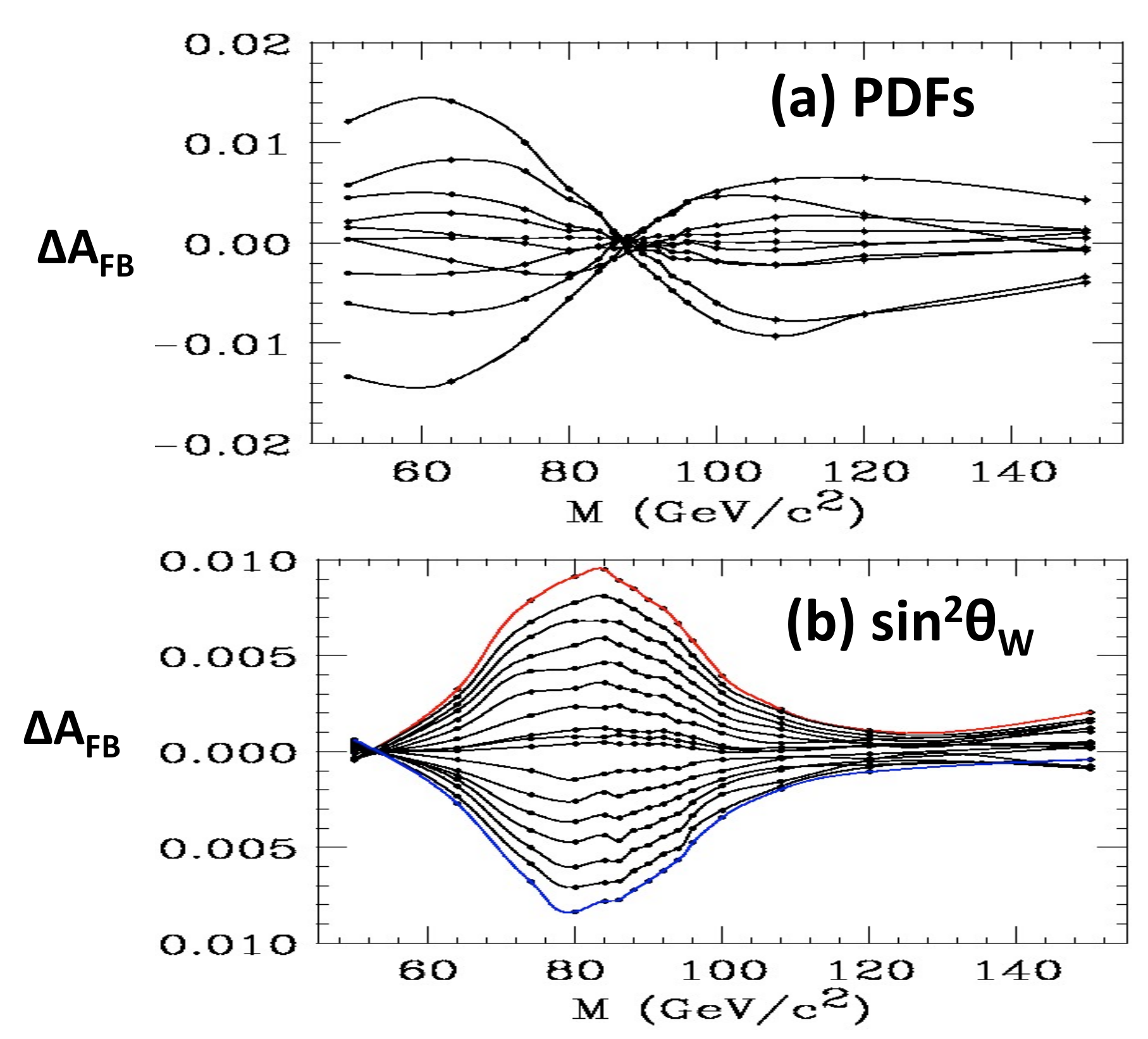}}
\caption{ Left panel: $A_{FB}(M)$ at the Tevatron
 for   $\bar{u}u$  (black), $\bar{d}d$ (red)
and the sum of the two  (blue). Right panel:  (a) The  difference between $A_{FB}(M)$ 
at the Tevatron 
for 10 NNPDF3.0 replicas and $A_{FB}(M)$ calculated
for  the  default NNPDF3.0. 
Here $\sin^2\theta_W$ is fixed at a value of 0.2244. (b) 
The  difference between 
$A_{FB}(M)$  for different values of $\sin^2\theta_W$ ranging
from 0.2220 (shown at the top in red)  to 0.2265 (shown on the
bottom in blue), and $A_{FB}(M)$ for $\sin^2\theta_W$=0.2244. Here,
the NNPDF3.0 default PDF is used.
}
\label{fig_1}
\end{figure}

%
For $\bar{p}p$ collisions, the direction of
the quark is predominately in the proton direction, and  the
direction of the antiquark is predominately in the antiproton
direction. Here,   most of the cross section
originates from the annihilation of quarks in the proton
with antiquarks in the antiproton. Therefore, 
$A_{FB}$ is measured under the assumption that the quarks
originate form the proton, and the antiquarks originate from the 
antiproton. 

The extraction of $\sin^2\theta_{eff}^{lept}$ from
$A_{FB}(M)$  at the Tevatron is sensitive to PDFs
for two reasons.  First, $A_{FB}(M)$ for up 
and  down type  quarks is different
as shown in Fig. \ref{fig_1}.  The asymmetry at the Tevatron originates
 primarily from up quarks and 
 is diluted  by the fraction of down quarks
in the proton because the  asymmetry
  for down quarks is much smaller.
In addition, there is a  small fraction of events for
which the annihilation is between sea antiquarks in the
proton with a sea quarks in the antiproton.
This also results in a  dilution  of  the
measured asymmetry. %
%
%

The  mass dependence  of $A_{FB}(M)$ depends on both  $\sin^2\theta_W$
 and   on PDFs    In the region of the $Z$ pole,   $A_{FB}(M)$ 
 is sensitive to the vector coupling, which depend on   $\sin^2\theta_W$.
 At higher and lower mass  $A_{FB}(M)$ is sensitive to the
 axial coupling and therefore insensitive to value of  $\sin^2\theta_W$.
  In contrast,  the magnitude of the  dilution of  $A_{FB}(M)$ depends on  the PDFs.
  The sensitivity to PDFs is largest in regions where   $A_{FB}(M)$ is large
  (i.e. away from the $Z$ pole). 

\begin{figure}[ht]
\includegraphics[width=7.5 cm, height=8.0 cm]{{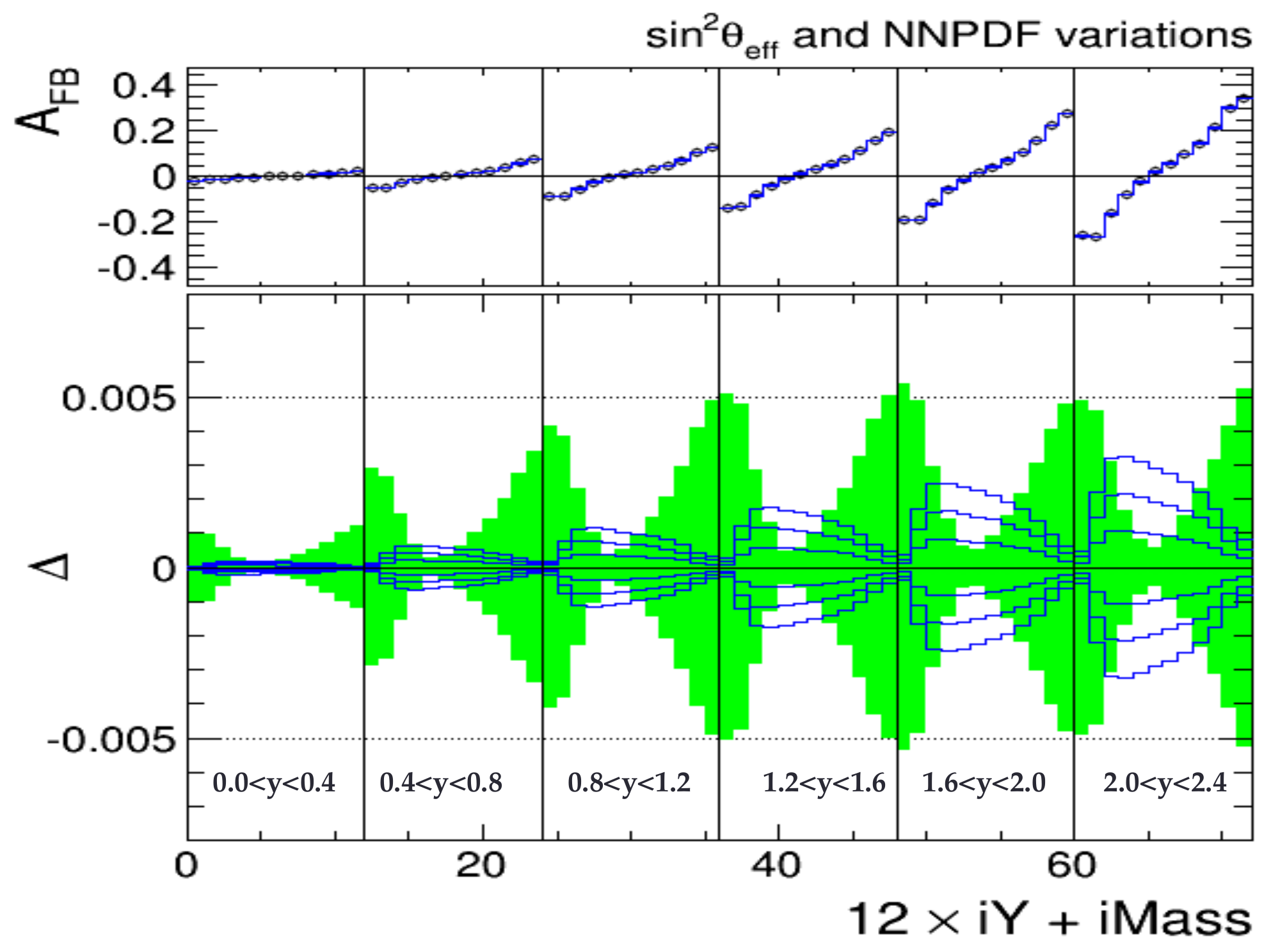}}
\includegraphics[width=7.5 cm, height=8.0 cm]{{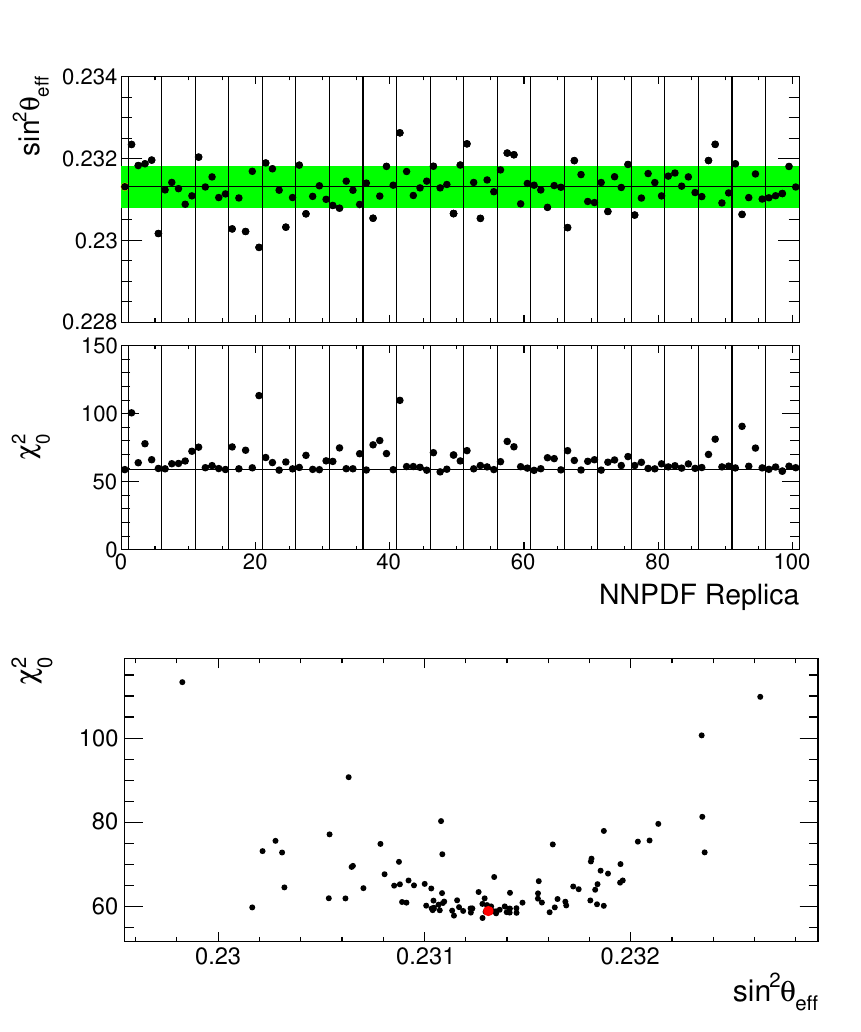}}
\caption{
LHC:  Left-Top panel- $A_{FB}$ at the LHC at $\sqrt{s}$=8 TeV for
six rapidity bins.
 The horizontal scale for each of  the six plots is the $\mu^+\mu^-$ invariant mass.
Left-Bottom panel :  The green bands span the  difference between $A_{FB}(M,y)$ 
 calculated for the 100 NNPDF3.0 replicas and $A_{FB}(M,y)$ calculated
for  the central  default NNPDF3.0   for the six $\mu^+\mu^-$ rapidity bins. The blue lines are
the  differences between 
$A_{FB}(M,y)$  calculated with  different values of $\sin^2\theta_{eff}$ 
and the values calculated with nominal  $\sin^2\theta_{eff}$=0.23120.
Right-Top panel.
Analysis of one of the 64  LHC pseudo experiments
 The top two panels show the  extracted $\sin^2\theta_{eff}$ and 
  corresponding  $\chi^{2}_{AFB}$  values  from fits
 to $A_{FB}(M,y)$ versus  replica
 number for  the 100 NNPDF3.0 replicas.
 The Right-Bottom panel shows the same results
 in the form of a scatter plot of $\chi^{2}_{AFB}$ values versus $\sin^2\theta_{eff}$
 for one pseudo experiment.  }
\label{fig_2}
\end{figure}

The right panel of Fig.~\ref{fig_1} shows the sensitivity of  $A_{FB}(M)$ 
at the Tevatron to PDFs.
Also  shown is  the sensitivity of  $A_{FB}(M)$ 
at the Tevatron to $\sin^2\theta_W$. 
 There 
is a large difference in the  $A_{FB}(M)$
predictions for PDF sets with different 
$\frac{d}{u} (x)$ and  $\frac{\bar{u}}{u}(x)$
in  regions where  $A_{FB}(M)$
is large and positive  (M$>$100 GeV).
 The changes in $A_{FB}(M)$ in  
regions where  $A_{FB}(M)$ is large and negative
(M$<$80 GeV) are  the opposite direction.
In contrast, different
values of  $\sin^2\theta_W$ change
$A_{FB}(M)$ primarily in the region  
near the $Z$ pole.  However, here the change
is in the same direction above and below the
$Z$ pole.  
Therefore, if we extract  $\sin^2\theta_W$ 
from $A_{FB}(M)$  data using different PDFs, 
PDFs with poor values of $\chi^2$ are less likely to be
correct.


The Left-Top  panel of Fig.~\ref{fig_2}  shows $A_{FB}(M,y)$ at the LHC at $\sqrt{s}$=8 TeV for
six rapidity bins (i=1 to 6) with average $y$ values of  0.2, 0.6, 1.0, 1.4, 1.8 and 2.2.
The horizontal scale for each of  the six plots is the $\mu^+\mu^-$ invariant mass. 
The calculations are done with the POWHEG  MC generator. The
version of POWHEG  that is used does not include electroweak radiative corrections.
Therefore, POWHEG requires an input value of  $\sin^2\theta_{eff}$ for the  calculation of $A_{FB}$. 
The green bands span  the  difference between $A_{FB}(M,y)$ 
calculated with the 10 NNPDF3.0 replicas and $A_{FB}(M,y)$ calculated
with the  default NNPDF3.0 PDF. 
 The blue lines are the  differences between 
$A_{FB}(M,y)$  calculated for several values of  $\sin^2\theta_{eff}$ 
and $A_{FB}(M,y)$ for the nominal  $\sin^2\theta_{eff}$=0.23120.
 For all of the blue lines,  $A_{FB}(M,y)$ is  calculated
with  the default  NNPDF3.0  PDF.

At the LHC, as for the Tevatron, the dependence of  $A_{FB}(M,y)$ on $\sin^2\theta_{eff}$
 and  on PDFs is different.    In the region of the $Z$ pole,   $A_{FB}(M,y)$ 
 is sensitive to the vector coupling, which are  functions of    $\sin^2\theta_{eff}$.
 At higher and lower mass  $A_{FB}(M,y)$ is sensitive to the
 axial coupling and therefore insensitive to value of  $\sin^2\theta_{eff}$.
 As is the case for the Tevatron, the magnitude of the  dilution of  $A_{FB}(M)$
  is larger in regions where the absolute value of   $A_{FB}(M)$ is large
  (i.e. away from the $Z$ pole). At the  LHC 
  the  dilution depends   on  both M and y.
 The  combined mass and rapidity dependence 
  of the dilution at the LHC provides  more stringent 
 constraints on PDFs than $A_{FB}(M)$ 
     measurements at the Tevatron.

%
The NNPDF3.0 PDF set is given in the form   
 of  N (e.g. 100 or 1000) 
replica PDFs. Each of the PDF replicas
has equal probability of being correct.  The
central value of any observable 
is the average of the values
extracted using each one of the N 
PDF replicas.  The PDF error is the RMS of the
values extracted using each of the N replicas. 

One advantage of the PDF replica method is that
constraints from new data can easily be incorporated
in any analysis by using different weights
for each replica.
Replicas for which the theory predictions 
are in agreement with the new data are given higher
weights, and replicas for which the predictions 
are in poor agreement are given lower weights.
The weights are derived from the $\chi^2$
values of the comparison between the new
data and theory prediction using 
each of the PDF  replicas.
The central value of any observable 
is then the {\it {\it weighted}}  average of the values
extracted using each one of the N 
PDF replicas.  The PDF error is the {\it {\it weighted}}  RMS of the
values extracted using each of the N replicas. 

The procedure of constraining a PDF set with
new data was initially proposed by   Giele and Keller\cite{GK}.
They proposed that  each of the N PDF replicas be {\it {\it weighted}} by $w_i$, and
the weights reduce the effective number of replicas\cite{Ball}  from N to $N_{eff}$.  Here
$$w_i=\frac {~ e^ {-\frac{1}{2}\chi^2_i}}{\frac {1}{N} 
\sum_{i=1}^{N}{~ e^ {-\frac{1}{2}\chi^2_i}}}; ~~~~~~~~N_{eff}=exp ( \frac {1}{N}  \sum_{i=1}^{N}{~ w_i~ ln(N/w_i)})$$
More recent discussions of the method can be found in references \cite{weights-MST,weights-web,GK1,GK2,Ball}.
The mass and rapidity dependence of
$A_{FB}$ can be  used to both provide additional constraints on
 PDFs and reduce the PDF error in measurements  of $ \sin^2 \theta_W$.
 
 \begin{table}[htb]
\caption {Values of  $\sin^2\theta_W$ 
 with statistical errors and  PDF errors 
 expected for a
 15 fb$^{-1}$ Drell-Yan $\mu^+\mu^-$ sample at the LHC
 (at 8 TeV).
 The pseudo data is generated  by 
 the POWHEG MC generator  with the  default  NNPDF3.0 PDF, and  
$\sin^2\theta_{eff}$=0.23120.
   The  PDF error for a standard  analysis is compared to the PDF
error for an analysis with both $\chi^{2}_{AFB}$ {\it weighting}
and $\chi^2_{AFB}+\chi^2_{Wasym}$ {\it weighting}. In addition, expected errors
for larger statistical samples are shown.}
    \begin{center}
\begin{tabular}{|c||c||c||c||}
\hline
                  input&	LHC~CMS~like	& LHC~CMS~like & LHC~CMS~like  \\
  	POWEG	& Pseudo-Exp. &  Pseudo-Exp.&  Pseudo-Exp.  \\
 Default& {{~15~fb$^{-1}$}} ~8~TeV &  {{~19~fb$^{-1}$}} ~8~TeV&  {{~200~fb$^{-1}$}} ~13~TeV    \\
{NNPDF3.0}  & $6.7M~(\mu^+\mu^-)$&  $15M~(\mu^+\mu^-, ~e^+e^-)$& $120M~(\mu^+\mu^-)$		\\
(261000)  &reconst.~events &	reconst.~events	& econst.~events\\ 
\hline\hline
$\sin^2\theta_{eff}$ statistical~error   	&	$\pm$0.00050&$\pm$0.00034	&$\pm$0.00011	\\
$\sin^2\theta_{eff}$ CT10~PDF~error  & $\pm$ 0.00080& & \\
\hline \hline
NNPDF3.0 Average   &	$N_{eff}=100$&&			\\
{PDF~error~RMS}   	&		$\pm$0.00051&&	\\
\hline
$\chi^2_{AFB}$ {\it weighting}		&	$N_{eff}=46$&	&			\\
{{\it {\it weighted}}~PDF~error RMS}		&		$\pm$0.00029	&&	\\
\hline
{$\chi^2_{AFB}$+$\chi^2_{Wsym}$~{\it weighting} }		&	$N_{eff}=21$&	&			\\
{{\it {\it weighted}}~PDF~error RMS}		&		$\pm$0.00026	&$\pm$ 0.00022& $\pm$0.00014	\\
\hline\hline
$\Delta\sin^2\theta_{eff}$  Stat+PDF &$\pm$0.00056 & $\pm$0.00040& $\pm$0.00018\\
$\Delta M_W$ indirect  Stat+PDF &$\pm$28~MeV &$\pm$ 20~MeV&$\pm$9 MeV\\
\hline\hline
\end{tabular}
\label{table_1}
    \end{center}
\end{table}
 For studies of  $A_{FB}(M,y)$ at the LHC we 
simulate Drell-Yan  $\mu^+\mu^-$ data for 64  pseudo experiments
for a  CMS like detector at $\sqrt{s}$=8 TeV.
The pseudo data is  generated 
using the  POWHEG  NLO MC
generator with the  default  NNPDF3.0 PDFs
and  $\sin^2\theta_{eff}$=0.23120.

 For each pseudo experiment,
 we generate a sample of 15.6  Million $\mu^+\mu^-$ events with $M_{\mu\mu} >$ ~50 GeV, which  corresponds to an integrated
luminosity of 15.0 fb$^{-1}$. 
 This is similar  to the  $\approx$19 fb$^{-1}$ of  integrated luminosity
collected by CMS and ATLAS  at 8 TeV. To this sample, we apply acceptance and transverse
momentum cuts which are similar to a CMS-like detector. We also smear the events with
a muon momentum resolution similar to a CMS-like detector. The final sample consists
 6.7M reconstructed $\mu^+\mu^-$ events.

The 8 TeV W asymmetry data at the LHC has not yet been incorporated
into the most recent PDF fits.  Therefore, 
in addition to  $A_{FB}(M,y)$, we also use the default NNPDF3.0 PDF to  generate pseudo data
for the  W decay muon asymmetry as a function of muon rapidity (for muon
transverse momentum PT$>$25 GeV). This simulates the  W asymmetry 
measurement at 8 TeV.

In the analysis of each of the   64 pseudo experiments  generated
with the default NNPDF3.0 PDF the  simulated
values of $A_{FB}(M,y)$  for each experiment are  compared
 to  $A_{FB}(M,y)$ templates. The templates are generated with the POWHEG MC for a range
 of values of $\sin^2\theta_{eff}$ for each of the 
    100 NNPDF3.0 PDF replica.  For each replica
 we extract the best fit value of  $\sin^2\theta_{eff}$, the corresponding
statistical error and the fit  $\chi^{2}_{AFB}$.  

In addition, we calculate $\chi^2_{Wasym}$ which is the $\chi^2$
for the agreement between the predictions for the W lepton decay
asymmetry and the W decay lepton asymmetry pseudo data at 8 TeV
for each  of the 100 PDF replicas.

The right panels of Fig. \ref{fig_2}  shows the results from one of the 
64 pseudo experiments at the LHC.  The top two  panels on the right
show the  extracted $\sin^2\theta_{eff}$ and 
  corresponding  $\chi^{2}_{AFB}$  values  from fits
 to $A_{FB}(M,y)$ versus  replica
 number for  the 100 NNPDF3.0 replicas.
 The bottom panel on the right  shows  the same results in the form of a 
   scatter plot of $\chi^{2}_{AFB}$ values versus $\sin^2\theta_{eff}$
   for one pseudo experiment.
  For each pseudo experiment we  find   the mean value and PDF error of  $\sin^2\theta_{eff}$ 
  from   the average and RMS of the $\sin^2\theta_{eff}$ extracted values using each of the  100 PDF replicas. 
   The average and RMS values are done in  three ways:
(1)      Using the standard  average and RMS of the  $\sin^2\theta_{eff}$ fit  values.
              This analysis results in a standard  PDF error of $\pm0.00051$  with 100 replicas.
(2)   Using the  $\chi^{2}_{AFB}$ values  of the fits to $A_{FB} (M,y)$  to form  a {\it {\it weighted}} average and 
     {\it {\it weighted}} RMS of the  $\sin^2\theta_{eff}$  values.
        This analysis results in a  PDF error of $\pm0.00029$  with 46 effective  replicas.
(3)  Using the combined  $\chi^{2}_{AFB}$+$\chi^2_{Wasym}$ 
        for the fits to Drell-Yan $A_{FB} (M,y)$ pseudo data  and the fits to the 
 W lepton decay asymmetry pseudo data   to form the {\it weighted} average and {\it weighted} RMS 
  of the  $\sin^2\theta_{eff}$  values.
   This analysis results in a  PDF error of $\pm0.00026$  with 21 effective  replicas.
 
Table~\ref{table_1} shows  values of  $\sin^2\theta_W$ 
 with statistical errors and  PDF errors 
 expected for a
 15 fb$^{-1}$ Drell-Yan $\mu^+\mu^-$ sample at the LHC
 (at 8 TeV).
 The pseudo data is generated  by 
 the POWHEG MC generator  with the  default  NNPDF3.0 PDF, and  
$\sin^2\theta_{eff}$=0.23120.
   The  PDF error for a standard  analysis is compared to the PDF
error for an analysis with both $\chi^{2}_{AFB}$ {\it weighting}
and $\chi^2_{AFB}+\chi^2_{Wasym}$ {\it weighting}. 
  As shown in Table~\ref{table_1},  the number of effective PDF replicas is reduced when we
  apply constraints from $\chi^{2}_{AFB}$ and   $\chi^2_{Wasym}$.  Therefore, the analysis
  will be  more robust if we start with  1000 PDF replicas. 
  
Also shown are the  expected errors for larger statistical samples. With larger
statistical samples, the PDF constraints are more stringent, and the PDF errors
are also reduced.  The  errors in this
   indirect measurement of the W mass are competitive with direct measurements. For
   example, with 200 fb$^{-1}$ at 13 TeV, the expected error in the indirect measurement of
   the W mass is $\pm$9 MeV.  Additional details and studies for both the Tevatron
   and LHC are give in ref.~\cite{bodek-afb}.

\end{document}